\documentclass{jfm}
\usepackage{graphicx}
\usepackage{natbib}
\usepackage{bm, amsmath}

\usepackage{url}		%
\let\realverbatim=\verbatim
\let\realendverbatim=\endverbatim
\renewcommand\verbatim{\par\addvspace{6pt plus 2pt minus 1pt}\realverbatim}
\renewcommand\endverbatim{\realendverbatim\addvspace{6pt plus 2pt minus 1pt}}
\makeatletter
\newcommand\verbsize{\@setfontsize\verbsize{10}\@xiipt}
\renewcommand\verbatim@font{\verbsize\normalfont\ttfamily}
\makeatother
\ifCUPmtlplainloaded \else
  \checkfont{eurm10}
  \iffontfound
    \IfFileExists{upmath.sty}
      {\typeout{^^JFound AMS Euler Roman fonts on the system,
                   using the 'upmath' package.^^J}%
       \usepackage{upmath}}
      {\typeout{^^JFound AMS Euler Roman fonts on the system, but you
                   don't seem to have the}%
       \typeout{'upmath' package installed. jfm.cls can take advantage
                 of these fonts,^^Jif you use 'upmath' package.^^J}%
       \providecommand\mathup[1]{##1}%
       \providecommand\mathbup[1]{##1}%
      }
  \else
    \providecommand\mathup[1]{#1}%
    \providecommand\mathbup[1]{#1}%
  \fi
\fi

\ifCUPmtlplainloaded \else
  \checkfont{msam10}
  \iffontfound
    \IfFileExists{amssymb.sty}
      {\typeout{^^JFound AMS Symbol fonts on the system, using the
                'amssymb' package.^^J}%
       \usepackage{amssymb}%
       \let\le=\leqslant  
       \let\ge=\geqslant  
      }{}
  \fi
\fi

\ifCUPmtlplainloaded \else
  \IfFileExists{amsbsy.sty}
    {\typeout{^^JFound the 'amsbsy' package on the system, using it.^^J}%
     \usepackage{amsbsy}}
    {\providecommand\boldsymbol[1]{\mbox{\boldmath $##1$}}}
\fi

\providecommand\bnabla{\boldsymbol{\nabla}}
\providecommand\bcdot{\boldsymbol{\cdot}}

\newsavebox{\astrutbox}
\sbox{\astrutbox}{\rule[-5pt]{0pt}{20pt}}

\newcommand\p{\ensuremath{\partial}}

\newcommand\rt{\rho_{\rm t}}
\newcommand\rb{\rho_{\rm b}}
\newcommand\etat{\eta_{\rm t}}
\newcommand\etab{\eta_{\rm b}}
\newcommand\e{{\rm e}}
\renewcommand\i{{\rm i}}

\newcommand\hphi{\hat \phi}
\newcommand\Nfl{N_{\rm F}}

\newcommand\Tv{T_{\rm v}}
\title[Simulation of 2D Faraday waves with phase-field modelling]
{Numerical simulation of two-dimensional Faraday waves with phase-field modelling}
\author[K. Takagi and T. Matsumoto]{Kentaro Takagi and Takeshi Matsumoto}%{A.\ns J.\ns W\ls O\ls O\ls L\ls L\ls A\ls T\ls T$^1$}

\affiliation{$^1$ Division of physics and astoronomy, guraduate school
of science, Kyoto University, Kitashirakawa Oiwaketyo Sakyoku Kyoto
606-8502 Japan}

\date{21 February 2012}
\pubyear{2011} % only needed when year of publication is not current year.
\volume{000}
\pagerange{\pageref{firstpage}--\pageref{lastpage}}
\doi{S002211200100456X}

\begin{document}

%\label{firstpage}
\maketitle

\begin{abstract}
A fully nonlinear numerical simulation
of two-dimensional Faraday waves between two incompressible
and immiscible fluids is performed by adopting
the phase-field method with the Cahn-Hilliard equation due
to Jacqmin (\textit{J. Comput. Phys.}, vol. 155, 1999, pp. 96-127).
Its validation is checked against the linear theory. 
In the nonlinear regime, qualitative comparison is made with an 
earlier vortex-sheet simulation of two dimensional Faraday waves by
Wright , Yon \& Pozrikidis (\textit{J. Fluid Mech.}, vol. 400, 2000, pp. 1-32).
The vorticity outside the interface region is studied in this comparison.
The period tripling state, which is observed in the quasi-two dimensional 
experiment by Jiang, Perlin \& Schultz (\textit{J. Fluid Mech.}, vol. 369, 1998, pp. 273-299), 
is successfully simulated with the present phase-field method.
\end{abstract}


\section{Introduction}
Faraday waves, which typically refer
to complex patterns of standing waves 
on a fluid surface in an oscillating container,
are among the classical problems of fluid
mechanics \citep{Faraday}. 
The phenomenon
has been a representative example of parametric instabilities \citep{Miles90}.
As is often the case with fluid mechanics, 
even classical phenomena are often not well understood in nonlinear regimes.
Indeed, continuing experiments on Faraday waves beyond the linear regimes 
reveal intriguing new features, which
include snake-like structures in drop-confined Faraday waves \citep{couder11},
a turbulent state mediated by defects of the pattern \citep{fineberg10}
and so-called oscillons \citep{fineberg00}.

These surprising findings would probably be
outside of the applicable range of the weakly nonlinear
theories, such as the one developed by \cite{ChenVignal99}, 
on selection of various patterns of Faraday waves.
Hence, to understand these phenomena,
fully nonlinear numerical simulations of the Faraday systems,
which can be complementary to laboratory experiments, play
an indispensable role as discussed in \cite{wagner05}.
Perhaps the first such simulation, in which the motions of both
the top and bottom fluids are simultaneously simulated, 
was performed recently by \cite{pjt}. This motivates our present study.

In these fully nonlinear simulations, the inevitable difficulties
are how to represent numerically the interface between the two
immiscible fluids, how to follow its motion and how to calculate 
its influence on the bulk fluids.
In \cite{pjt},
the interface is modelled by triangular elements whose apices are
advected vertically and data on the interface elements are copied
to the Eulerian grids in the bulk of the fluids according to the
recipe of the immersed boundary method \citep{Peskin}.
Since numerical modelling of the interface involves various assumptions,
its validation is required 
against laboratory experimental data or other data independent of the modelling.
In the linear regime, the analytical result of 
the linear theory of Faraday waves due to \cite{kt94} provides
reliable data for comparison, as used in \cite{pjt}.
In nonlinear regimes, comparison with experimental data is crucial. 
Remarkably, the simulation result by \cite{pjt} 
is in perfect agreement with the experimental data in the nonlinear 
regime conducted by \cite{wagner05}, where the top fluid's motion cannot be neglected.

The further task of such validated numerical simulations is to
investigate data not easily accessible in experiments, such as nonlinear
energy transfers between modes.
For this purpose, in our opinion, it is important to have 
at least two validated numerical simulations with independent 
interface modelling and make sure that these simulations give
consistent results.

The aim of this study is to develop another nonlinear
simulation method of Faraday waves with an alternative interface
model to that devised by \cite{pjt}. 
We here apply to the Faraday wave problem the phase-field modelling 
with the Cahn-Hilliard equation for binary fluids 
\citep{Jacqmin}.
The phase-field method, like front-tracking methods, volume-of-fluid methods and level-set
methods, easily allows situations where the interface becomes a multivalued function of the
horizontal coordinates.
See, for example, \cite{cmmv09} for recent application of the
phase-field method to Rayleigh-Taylor 
instability and to other flows in the references therein. 
To our knowledge, the method hat not been applied to the Faraday wave problem.

%%%%%%%%%%%%%%%%%%%%%%%%%%%%%%%%%%%%%%%%%%%%%%%%%%%%%%%%%%%%%%
In this paper we focus on Faraday waves in two spatial
dimensions (2D) to explore the capabilities of the phase-field 
method in a simple setting.
Earlier numerical studies of 2D Faraday
waves include \cite{chenwu}, \cite{murakami} and \cite{ubal}, 
where the fluid dynamical equations for the bottom fluid are solved 
but the top fluid's motion is neglected in contrast to the present 
simulation.
Another numerical approach using different formalisms (the boundary 
integral and the vortex sheet) is explored by \cite{Poz00}.

Here we proceed as follows. 
In the linear regime, we compare quantitatively the phase-field
simulation with the linear theory by \cite{kt94} for validation 
as in \cite{pjt}. 
In the nonlinear regime, unfortunately suitable experimental data
in 2D for comparison with our simulation are not available.
However, we qualitatively make comparison with the simulation of 
2D Faraday waves with the vortex-sheet formulation by \cite{Poz00} in 
the regime of plume formation, where the interface becomes 
a multivalued function. 
Finally, we present simulation results concerning the period tripling state,
where the oscillation period of the Faraday waves becomes three times 
the basic period. This state is found experimentally by
\cite{jian98} in quasi-two dimensional Faraday waves,
by \cite{dh08} in axisymmetric three-dimensional Faraday waves,
and numerically also by \cite{Poz00}.
Its underlying physics is not yet understood.
%%%%%%%%%%%%%%%%%%%%%%%%%%%%%%%%%%%%%%%%%%%%%%%%%%%%%%%%%%%%%%w

The organization of the paper is the following.
In \S 2, we describe the basic fluid dynamical
the equations and our phase-field model of the Faraday waves.
Our numerical method for solving the equations is explained in \S 2.
We then make a quantitative comparison of its
simulation results with the linear analysis in \S 3.
In \S 4, we present results of the phase-field simulation
in nonlinear regimes.  
The first simulation shows plume formation, in which the interface
overturns. The second one concerns
the period tripling state.
Concluding remarks are made in the last \S 5.

\section{Equations and numerical method}
\subsection{Equations}
We consider Faraday waves between two immiscible fluids in two
spatial dimensions.
The fluid dynamical equations  are the incompressible
Navier-Stokes equations
\begin{eqnarray}
 \rho \left[\frac{\p \bm{u}}{\p t}
  + (\bm{u} \bcdot \bnabla) \bm{u} \right]&=&
  - \bnabla p + \rho \bm{G} + 
  \bnabla \bcdot \eta (\bnabla \bm{u} 
  + \bnabla \bm{u}^{T}) + \bm{s},\label{ns}\\
 \bnabla \bcdot \bm{u} &=& 0, \label{incomp}
\end{eqnarray}
where $p$ is the pressure and $\bm{u}$ is the velocity.
Here $\bm{s}$ is the surface tension force, which will be
discussed below. 
The density $\rho$ and viscosity $\eta$ are distinct constants
for each fluid, which are denoted as $\rt, \rb$ and $\etat, \etab$ respectively,
for the top and bottom fluids ($\rt < \rb$).
Finally, $\bm{G}$ is the gravitational acceleration, incorporating the 
vibration forcing with amplitude $a$ and angular frequency $\omega$, which is
written as
\begin{eqnarray}
 \bm{G} = (-g + a \cos\omega t) \bm{e}_z.
\label{G}  
\end{eqnarray}
Here in two dimensions, $\bm{e}_z = (0,\, 1)$.
The temporal period of the vibration forcing is denoted 
by $\Tv =2\pi/\omega$, which will be used often in what follows.

\begin{figure}
 \centerline{\includegraphics[height=2.5in,width=4in]{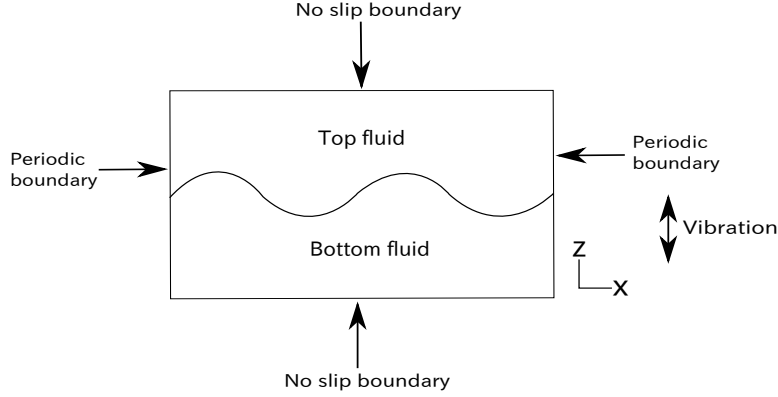}}
 \caption{Faraday wave setting in two-dimensional space.}
 \label{f:setting}
\end{figure}

The equations are solved with the boundary conditions of the
Faraday setting as depicted in figure \ref{f:setting}.
In the horizontal direction (coordinate $x$),
we assume periodic boundary conditions with length $L_x$. 
For the vertical direction (coordinate $z$), no-slip boundary
conditions at the boundaries $z = 0, L_z$ 
are assumed, that is, $\bm{u}(x, 0, t) = \bm{u}(x, L_z, t) = 0$.
The interface position between the top and bottom fluids, 
denoted as $z = \zeta(x, t)$, obeys the kinematic boundary 
condition.
In terms of this $\zeta(x, t)$, the density $\rho$
and viscosity $\eta$ in (\ref{ns}) are written as function of $z$:
\begin{eqnarray}
 (\rho,\, \eta) = 
  \begin{cases}
   (\rt,\, \etat)  & z > \zeta(x, t), \\			
   (\rb,\, \etab)  & z \le \zeta(x, t). 
  \end{cases}
\end{eqnarray}
In this sharp interface formulation, the density and viscosity vary 
discontinuously as a function of $z$, which makes numerical simulations in this form
very difficult in practice. One standard way to overcome this in numerical
simulations is to model the sharp interface by a diffuse one  
\citep{amw98}. The phase-field modelling is one such
diffuse-interface method, which we will explain below.

We adapt the phase-field modelling of binary fluids \citep{Jacqmin}
to Faraday waves. 
The top and bottom fluids are here indicated by the phase
variable values
$\phi = 1$ and $\phi = -1$, respectively.
The interface is modelled as the region where $\phi$ changes continuously
from $-1$ to $1$. The phase evolves to minimize the free energy 
functional of the phase $\phi$:
\begin{eqnarray}
 F[\phi(\bm{x}, t)] = 
\frac{\Lambda}{2} 
  \int \left[
  \frac{1}{2\epsilon^2}(\phi^2 - 1)^2
  +
  |\bnabla \phi|^2
\right]
  d\bm{x}.
\end{eqnarray}
Accordingly, the equation of the phase variable is
the Cahn-Hilliard equation with the advection term
\begin{eqnarray}
&& \frac{\p \phi}{\p t} + \bm{u} \bcdot \bnabla \phi
  = \gamma \bnabla^2 \mu, \label{ch} \\
&& \mu = \frac{\delta F}{\delta \phi}
     = \Lambda \left( - \bnabla^2 \phi
	       + \frac{\phi^3 - \phi}{\epsilon^2}\right).
\label{mu}
\end{eqnarray}
where $\mu$ is the chemical potential, $\Lambda$ is the magnitude of the
free energy and $\gamma$ is the mobility. We assume that the mobility
$\gamma$ is constant. The parameter $\epsilon$
controls the thickness of the smoothed interface.
These $\gamma$ and $\epsilon$ are adjustable parameters of the
phase-field modelling. How to determine
them is discussed at the end of section \ref{s:linear}.
In terms of  $\phi$, the
density and viscosity are expressed as
\begin{eqnarray}
 \rho = \frac{\rt + \rb}{2}
  + \frac{\rt - \rb}{2} \phi, \qquad
 \eta = \frac{\etat + \etab}{2}
  + \frac{\etat - \etab}{2}\phi.
\label{rhoeta}  
\end{eqnarray}

In the framework of the phase-field modelling, 
the surface tension force
written as $\bm{s}$ in (\ref{ns}) is given by
\begin{eqnarray}
 \bm{s} =  \mu  \bnabla \phi
\label{s}  
\end{eqnarray}
\citep{Jacqmin, cmmv09}.
The surface tension $\sigma$ used in the sharp interface model
is related to the phase-field simulation parameters as
\begin{eqnarray}
 \sigma = \frac{2\sqrt{2}}{3}\frac{\Lambda}{\epsilon}
  \label{sigma}
\end{eqnarray}
for planar interfaces. 
In the following we assume that
the correspondence (\ref{sigma}) is valid for non-planar interfaces.
We use the following boundary conditions of the phase variable 
on the top and bottom walls $z = 0, L_z$: 
\begin{eqnarray}
\bm{e}_z \bcdot \bnabla \phi = 0, \qquad \bm{e}_z \bcdot \bnabla \mu = 0
 \label{bphi}
\end{eqnarray}
\citep{Jacqmin}. 
In the horizontal direction, the boundary condition is periodic.
For further details of the derivation of the phase-field model,  
we refer the reader to \cite{Jacqmin}.

To summarise this subsection, the equations to be solved numerically
are the incompressible Navier-Stokes equations (\ref{ns}) and
(\ref{incomp}) and the Cahn-Hilliard equation (\ref{ch}). 
The boundary conditions on the velocity are periodic 
in the horizontal direction and no-slip on the top and bottom walls
in the two-dimensional space. 
For the phase variable, the boundary conditions are periodic in 
the $x$-direction and (\ref{bphi}) in the $z$-direction.

\subsection{Numerical method}
\label{ss:nm}
We start with the discretization of the Cahn-Hilliard equation.
The computational mesh is a standard staggered arrangement 
as in \cite{pjt}.
We Fourier-expand the phase variable $\phi$ in the periodic $x$-direction as
\begin{eqnarray}
 \phi(x, z, t) = \sum_{j} \hat{\phi}(k, z, t) ~\e^{\i k_jx}.
\end{eqnarray}
The wavenumber is given by $k_j = (2\pi / L_x) j \quad (j = -N_x/2 +
1,..., N_x / 2)$
, where $N_x$ is the number of grid points in 
the $x$-coordinate.
Then the Cahn-Hilliard equation in the $(k, z)$ space becomes 
\begin{eqnarray}
 \frac{\p \hphi}{\p t}
 + \widehat{(\bm{u}\!\bcdot\!\bnabla)\phi}
 = 
\gamma \Lambda
 \left\{
  -\left(-k^2 + \frac{\p^2}{\p z^2}\right)^2 \hphi
  -\frac{1}{\epsilon^2}
     \left[
          \left(-k^2 + \frac{\p^2}{\p z^2} \right)
          \left(\hphi - \widehat{\phi^3}\right)
   \right]
 \right\},
\label{ch-hat} 
\end{eqnarray}
where the cubic term $\widehat{\phi^3}$ is the Fourier
transform of the product $\phi^3$ calculated in the physical space $(x, z)$.
No aliasing error is removed in this calculation of $\widehat{\phi^3}$.

The advection term $\widehat{(\bm{u}\bcdot\bnabla)\phi}$ is calculated
as follows. First it is calculated in the physical space
by using an elementary discretization in the staggered mesh as
\begin{eqnarray}
&&\left[
 (\bm{u}\bcdot\bnabla)\phi 
\right]_{l,\, m}
 = \left(
    u\frac{\p\phi}{\p x}
  + w\frac{\p\phi}{\p z}
   \right)_{l,\, m}  
\nonumber\\ 
&&= \frac{   u_{l - 1/2,\, m}       +    u_{l + 1/2,\, m      }}{2}
    \,\,\frac{\phi_{l + 1,\,   m}       - \phi_{l - 1, \,  m      }}{2 \Delta x}
  + \frac{   w_{l,\,       m - 1/2} +    w_{l,    \,   m + 1/2}}{2}
    \,\,\frac{\phi_{l,\,       m + 1}   - \phi_{l,    \,   m - 1   }}{2 \Delta
    z}\nonumber\\.
\label{ch-ad-re}
\end{eqnarray}
In this notation $u_{l - 1/2,\, m}$ denotes the velocity on the cell
face $(x_{l - 1/2}, z_m)$ 
and $\phi_{l, m}$ denote the phase variable on the cell centre $(x_l, z_m)$.
The indices run as $l = 1, \cdots, N_x$ and $m = 1, \cdots, N_z$ ($N_z$ is
the number of grid points on the $z$-coordinates).
In the denominator $\Delta x$ and $\Delta z$ are the grid spacings of 
the $x$ and $z$ coordinates. The Fourier transform of (\ref{ch-ad-re})
gives $\widehat{(\bm{u}\bcdot\bnabla)\phi}$. Again no aliasing error is
removed.

Equation (\ref{ch-hat}) is discretized in time by 
using the semi-implicit stabilized scheme due to \cite{Eyre}
as 
\begin{eqnarray}
&& \frac{\hphi^{(n + 1)} - \hphi^{(n)}}{\Delta t}
 + \widehat{(\bm{u}^{(n)} \bcdot \bnabla) \phi^{(n)}} 
\nonumber\\
&&= \gamma \Lambda \left\{
 - \left(-k^2 + \frac{\p^2}{\p z^2} \right)^2 \hphi^{(n+1)}
 - \frac{1}{\epsilon^2}
 \left[
  \left(-k^2 + \frac{\p^2}{\p z^2}\right)
  \left(
   3\hphi^{(n)} - 2\hphi^{(n+1)} - \widehat{(\phi^{(n)})^3}
  \right)
  \right]
 \right\}.
\nonumber\\
 \label{plm}
\end{eqnarray}
Here $\hat{\phi}^{(i)}$ denotes the data at the $i$-th time step.
On the right hand side, the fourth-order derivative term is 
treated fully implicitly. 
The second-order term is handled semi-implicitly, which is
split into two terms: one involves $3\hphi^{(n)}$ and the 
other $2\hphi^{(n + 1)}$ as proposed by \cite{Eyre}.
The $z$-derivatives in (\ref{plm})
are evaluated as
\begin{eqnarray}
 \left(\frac{\p^2 \hphi}{\p z^2}\right)_{j,m} 
  &=& \frac{1}{\Delta z^2}\left(
     \hphi_{j, m + 1}
  -2 \hphi_{j, m}
  +  \hphi_{j, m - 1}
       \right),\\
 \left(\frac{\p^4 \hphi}{\p z^4}\right)_{j,m} 
&=& \frac{1}{\Delta z^4}\left(
     \hphi_{j, m + 2}
  -4 \hphi_{j, m + 1}
  +6 \hphi_{j, m}
  -4 \hphi_{j, m - 1}
  +  \hphi_{j, m - 2}
\right),
\end{eqnarray}
where $\hphi_{j, m}$ denotes $\hphi(k_j, m \Delta z)$.
We use the bi-conjugate gradient stabilized (BiCGSTAB) method \citep{Saad}
to solve the equations (\ref{plm}).
The boundary conditions of $\phi$ in the $z$-direction are 
given in (\ref{bphi}).

Now we turn to discretization of the incompressible Navier-Stokes 
equations (\ref{ns}) and (\ref{incomp}).
We follow mostly the discretization method proposed by 
\cite{pjt} except for the surface tension force.
In order to describe how we calculate the surface tension force in the
present phase-field modelling, 
we repeat here some of the equations given in \S 3.3 of \cite{pjt}.
The method starts with the intermediate velocity $\tilde{\bm u}$
from $\bm{u}^{(n)}$ at the $n$-th time step,
\begin{eqnarray}
 \frac{\tilde{\bm u} - \bm{u}^{(n)}}{\Delta t}
  = - (\bm{u}^{(n)}\bcdot\bnabla) \bm{u}^{(n)} 
    + \frac{1}{\rho^{(n+1)}}\bnabla \bcdot \eta^{(n+1)}
    (\bnabla \tilde{\bm{u}} + \bnabla \tilde{\bm{u}}^{T}),
\label{utilde}    
\end{eqnarray}
where $\rho^{(n)}$ and $\eta^{(n)}$ are here determined by
the phase $\phi^{(n)}$ as (\ref{rhoeta}). 
Yet another intermediate velocity $\bm{u}^*$ involves the gravity
and the vibration forces $\bm{G}$ and the surface tension force $\bm{s}$:
\begin{eqnarray}
 \frac{\bm{u}^* - \tilde{\bm{u}}}{\Delta t}
  = \bm{G}^{(n+1)}
  + \frac{\bm{s}^{(n+1)}}{\rho^{(n+1)}} 
  - \frac{1}{\rho^{(n+1)}} \bnabla p^{(n)}.
\end{eqnarray}
Since this $\bm{s}$ is represented as (\ref{s}) in the phase-field method,
we discretize each component first as 
\begin{eqnarray}
 \left( s_x \right)^{(n)}_{l,m}
  = &&\mu_{l,m}^{(n)} \frac{\phi^{(n)}_{l + 1, m} - \phi^{(n)}_{l - 1,
  m}}{2 \Delta x}, \quad
 \left( s_z \right)^{(n)}_{l,m}
 =  \mu_{l,m}^{(n)} \frac{\phi^{(n)}_{l, m + 1} - \phi^{(n)}_{l, m - 1}}{2 \Delta z}.
\end{eqnarray}
Then, the vectors $\bm s$ on the staggered grids are 
\begin{eqnarray}
 \left( s_x \right)^{(n)}_{l+1/2,m}
  = &&  \frac{\left( s_x \right)^{(n)}_{l+1,m} +  \left( s_x \right)^{(n)}_{l,m}}{2}, \quad
 \left( s_z \right)^{(n)}_{l,m + 1/2}
 =  \frac{\left( s_z \right)^{(n)}_{l,m + 1} +  \left( s_z \right)^{(n)}_{l,m}}{2}.
\end{eqnarray}
Here the chemical potential (\ref{mu}) is discretized as
\begin{eqnarray} 
 \mu^{(n)}_{l,m}
  &=& \Lambda \left[
    - \frac{\phi^{(n)}_{l + 1, m} - 2 \phi^{(n)}_{l,m} + \phi^{(n)}_{l - 1,m}}{\Delta  x^2}
    - \frac{\phi^{(n)}_{l, m + 1} - 2 \phi^{(n)}_{l,m} + \phi^{(n)}_{l,m - 1}}{\Delta  z^2}
    + \frac{(\phi^{(n)}_{l,m})^3 - \phi^{(n)}_{l,m}}{\epsilon^2}
\right].
\nonumber\\
\end{eqnarray}
Finally, the velocity at the $(n + 1)$-th time step is obtained by
\begin{eqnarray}
 \frac{\bm{u}^{(n+1)} - \bm{u}^{*}}{\Delta t}
  = -\frac{1}{\rho^{(n+1)}} \bnabla
  \left(p^{(n+1)} - p^{(n)}\right),
\end{eqnarray}
where the pressure $p^{(n + 1)}$ is determined,
by demanding  $\bnabla \bcdot \bm{u}^{(n+1)}=0$,  as
\begin{eqnarray}
 \frac{\bnabla \bcdot \bm{u}^*}{\Delta t}
  =  \bnabla \bcdot \frac{1}{\rho^{(n+1)}}
  \bnabla \left(
    p^{(n+1)} - p^{(n)}
  \right).
  \label{p}
\end{eqnarray}
Again the BiCGSTAB method is employed to solve (\ref{utilde}) and
(\ref{p}). The boundary conditions for the velocity and the intermediate
one $\tilde{\bm u}$ are the same as in \cite{pjt}, 
namely, $\tilde{\bm{u}}(x, z = 0 \,{\rm or}\, L_z, t) = 0$,
$\bm{e}_z \bcdot \bm{u}^{(n + 1)}(x, z = 0 \,{\rm or}\, L_z, t) = 0$ and
$\bm{e}_z \bcdot \nabla[p^{(n + 1)}(x, z = 0\, {\rm or}\, L_z, t) - p^{(n)}(x, z =
0 \,{\rm or}\, L_z, t)] = 0$. 
As proposed by \cite{pjt}, in solving the equations (\ref{utilde}), 
we use the latest updated velocity component of $\tilde{\bm u}$ to 
compute the other component. In this process, the order (which component of 
$\tilde{\bm u}$ is computed first and last) is also interchanged
at each time step to ensure symmetry.

\section{Comparison in the linear regime}
\label{s:linear}
We compare our simulation result with the linear theory
of \cite{kt94}. We follow here again the validation method 
proposed by \cite{pjt}.

In the linear theory elaborated by \cite{kt94}, 
the Faraday wave problem is formulated as a finite-depth, viscous 
binary fluid system with a sharp interface representation. 
This provides the critical value of the vibration forcing
amplitude $a_c$ as eigenvalues for a given perturbation 
$\e^{\i kx}$.
The Floquet modes $f_q(k)$ 
of the interface position $z = \zeta(x, t)$ can be calculated 
as eigenvectors.
In other words, the spatio-temporal variation reads\citep{kt94}
\begin{eqnarray}
 \zeta(x, t) \propto
  \e^{\i k x} ~\e^{(\beta + \i \alpha \omega) t}
  \sum_{q = -\Nfl}^{\Nfl} f_q(k) ~\e^{\i q \omega t}.
\label{floquet}
\end{eqnarray}

Recall that $\omega$ is the angular frequency
of the vibration forcing (\ref{G}).
Here the number of Floquet modes $\Nfl$ is infinite
in theory but it is finite for numerical calculations.
The exponent $\beta + \i \alpha \omega$ is 
the Floquet exponent. 
For the critical perturbations $\beta$ is zero.
The harmonic $\alpha = 0$ and subharmonic $\alpha = 1/2$ cases
are considered here as in \cite{pjt}.

\begin{figure}
 \centerline{\includegraphics[height=3in,width=4in]{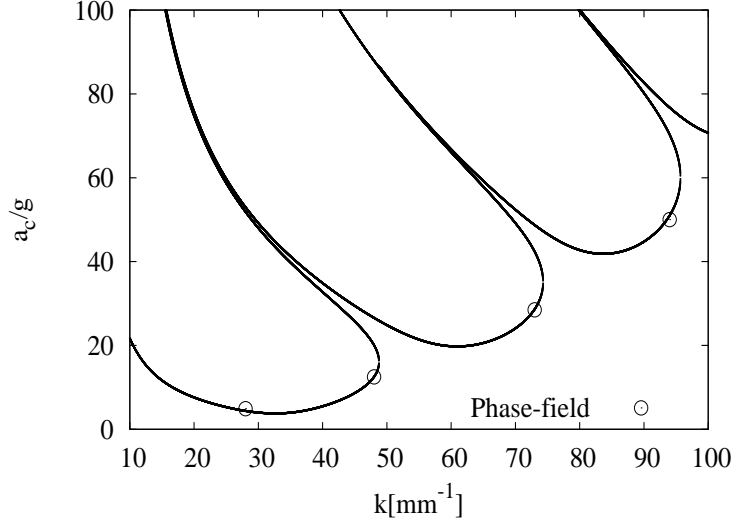}}
 \caption{Critical curve of the vibration amplitude from the linear
 theory \citep{kt94}. Points are the critical values determined with the phase-field simulation.}
 \label{f:ac}
\end{figure}
\begin{figure}
 \centerline{\includegraphics[height=2in,width=3in]{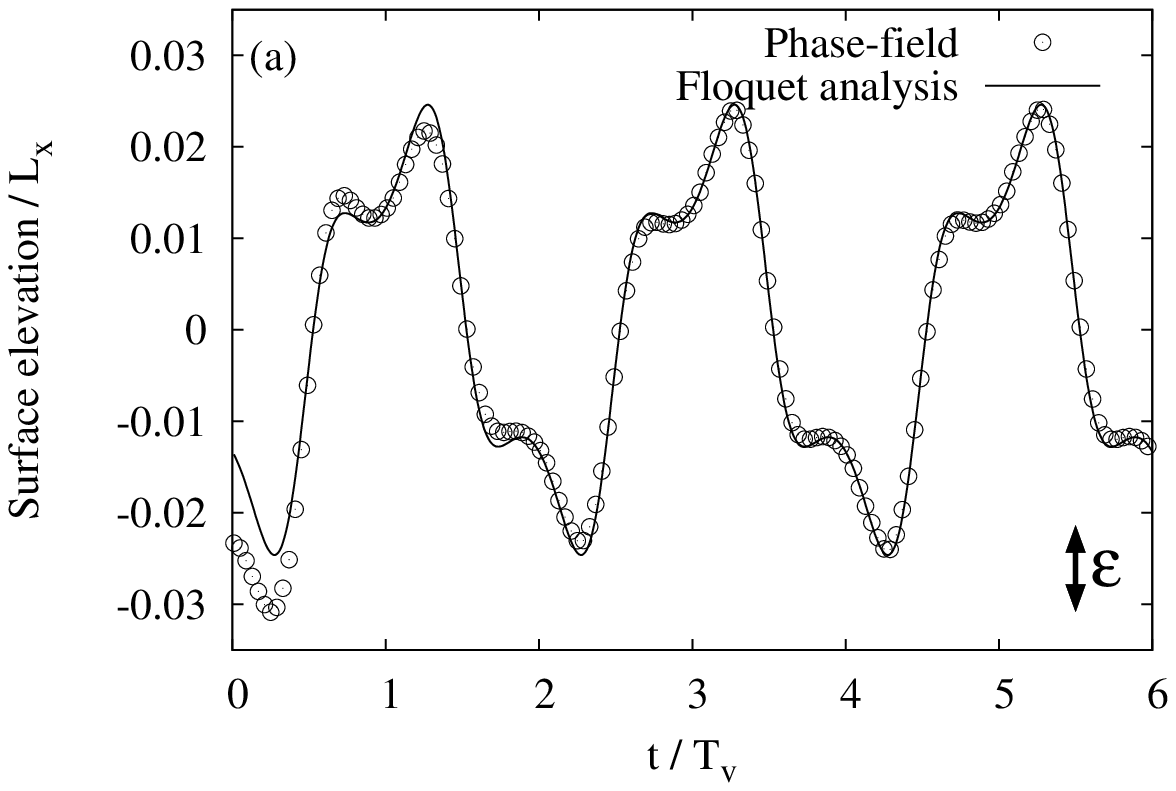}}
 \centerline{\includegraphics[height=2in,width=3in]{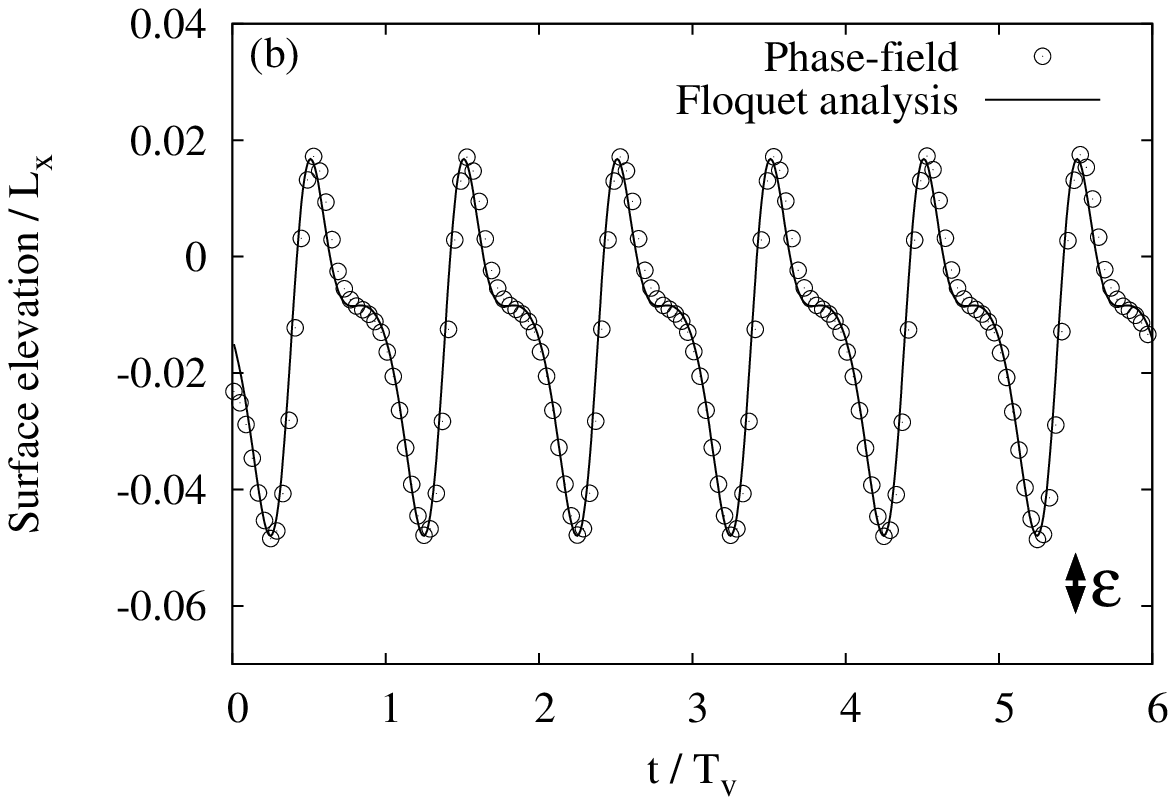}}
 \centerline{\includegraphics[height=2in,width=3in]{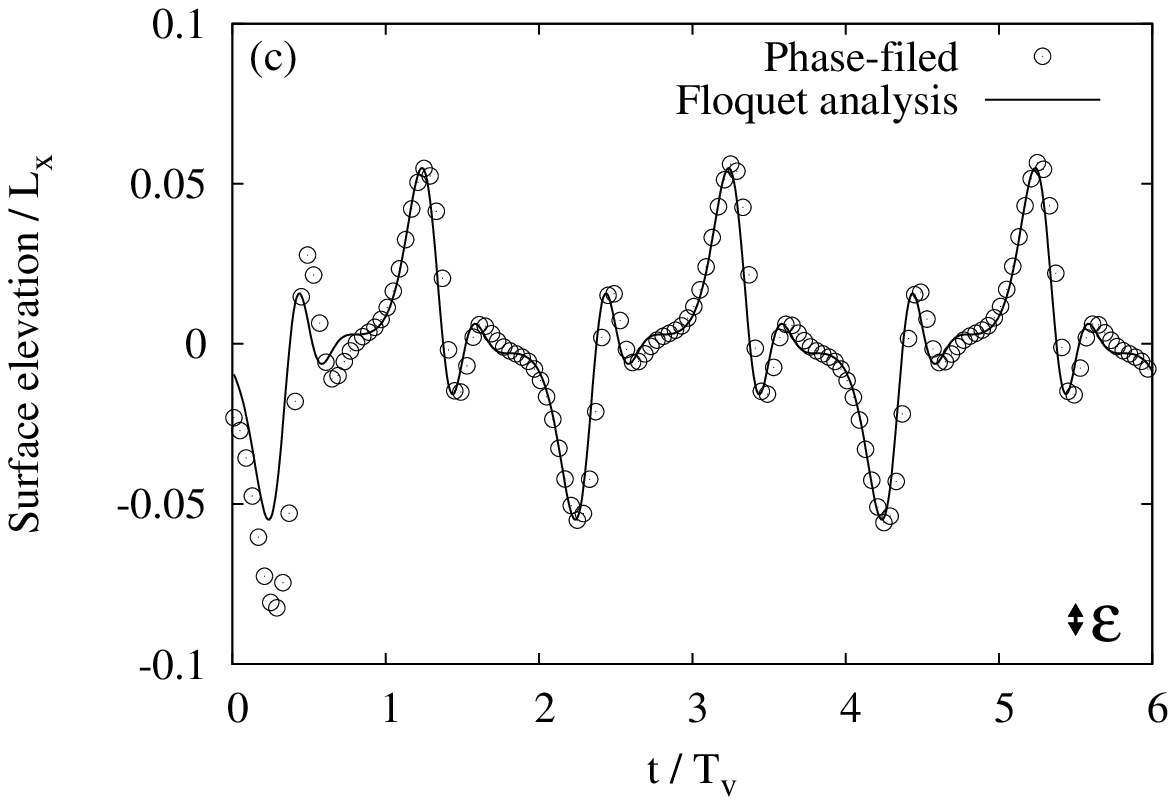}}
 \caption{Time variations of the interface position 
 at $x = L_x / 2$
 of the phase-field simulation (circles) and 
 the Floquet analysis result (curves). The interface position is
 defined as the phase null points $\phi = 0$.
 The thick arrows indicate the size of $\epsilon$, which is the length
 scale of the diffuse interface.
 (a) Perturbation wavenumber $k = 48.0$ mm$^{-1}$ (subharmonic case);
 (b) $k = 73.0$ mm$^{-1}$ (harmonic case); 
 (c) $k = 94.0$ mm$^{-1}$ (subharmonic case).}
 \label{f:floquet}
\end{figure}

We numerically solve this eigenvalue problem  
(for the precise form of the matrices, see \cite{kt94}).
The Mathematica script that we use here is available at
http://www.kyoryu.scphys.kyoto-u.ac.jp/\textasciitilde takeshi/kt94.
The parameters here are the same as in \cite{pjt} in their
validation with the linear theory. The parameter values are
$N_F = 10$, 
$\rho_{\rm b} = 5.19933 \times 10^{2}$ kg m$^{-3}$,
$\rho_{\rm t} = 4.15667 \times 10^{2}$ kg m$^{-3}$,
$\eta_{\rm b} = 3.908   \times 10^{-5}$ Pa s,
$\eta_{\rm t} = 3.124   \times 10^{-5}$ Pa s,
$\sigma       = 2.181   \times 10^{-6}$ N m$^{-1}$,
$g            = 9.8066                $ m s$^{-2}$,
$L_z          =  2.31 \times 10^{-4}$ m
and $\omega = 2\pi \times 10^2$ s$^{-1}$. 
The unperturbed interface is in the centre of the container.
The critical vibration forcing amplitude $a_c$ from the linear
theory as a function of the perturbation wavenumber $k$ 
is plotted as a curve in figure \ref{f:ac}. The Floquet modes
$f_q$ are used later in figure \ref{f:floquet}.

With the phase-field simulation, critical values $a_c$ 
for four different $k$'s are determined, which are denoted
as points in figure \ref{f:ac}. Refer also to table
\ref{t:ac} for a precise comparison with the values from the 
linear theory calculation. 
We see that agreement between the two is satisfactory.

The way to estimate $a_c$ in the phase-field simulation is as
follows. 
(i) We consider the sharp interface location $z = \zeta(x, t)$
in the linear analysis to be the null point $\phi = 0$ in the phase-field 
modelling. Numerically, such null points are calculated using the linear 
interpolation of $\phi$ data on the grid points (we use this
correspondence throughout this paper).
(ii) For a given acceleration amplitude $a$, we monitor the oscillating interface
position at $x = L_x / 2$ as a function of time. We perform simulations
by changing $a$ from the theoretical $a_c$ and estimate the critical
value at which the temporal oscillation neither decays nor grows.
More precisely, we take the absolute relative difference between the two
peak values of the interface positions around $t = 2\Tv$ and $4\Tv$.
If the difference is smaller than $10^{-3}$, we regard this $a$ as the
critical acceleration of the phase-field simulation.
The largest error shown in table \ref{t:ac} (the case of $k = 28.0$)
would be due to a relatively large $\Delta x = L_x/ N_x$.

Here we set the parameters of the phase-field simulation to be
$\epsilon = \Delta z = 1.80 \times 10^{-6}$ m, 
$\gamma = 6.33 \times 10^{-8} \,{\rm m^3 kg^{-1}s}$,
$\Lambda = 3\sigma / (2\sqrt{2} \epsilon)$ (see (\ref{sigma})) and $\Delta t = 2.5 \times 10^{-6}$ for all four cases.
The initial perturbed interface is written in terms of the phase variable as
\begin{eqnarray}
 \phi = \tanh\left\{\frac{z - [L_z/2 - b \cos (kx)]}{\sqrt{2}
	      \epsilon}\right\},
\label{e:tanh}
\end{eqnarray}
where $b$ corresponds to the perturbation amplitude of the interface
position. Here we take $b = 3 \Delta z = 3 \epsilon$.
We recall that in the unperturbed state the Cahn-Hilliard equation 
has the stationary solution 
$\phi(z) = \tanh[(z - L_z/2)/(\sqrt{2} \epsilon)]$ (see,
e.g. \cite{Jacqmin}). The initial velocity in the phase-field
simulation is set to zero.
\begin{table}
 \begin{center}
   \begin{tabular}{@{}ccccc@{}}
    $k\,({\rm mm}^{-1})$ & $L_x({\rm mm})$ &$a_c / g$ (theory) & $a_c / g$ (simu.) & \multicolumn{1}{c@{}}{error$\,(\%)$}\\[3pt]
    28.0                 & 0.224           & 4.37               & 4.65	   & 6.0\\  
    48.0                 & 0.131           & 12.5              & 12.5              & 0.0\\
    73.0                 & 0.0861          & 28.5              & 28.4              & 0.49\\
    94.0                 & 0.0668          & 51.0              & 50.0              & 2.0\\
   \end{tabular}
 \end{center}
 \caption{Critical amplitude values $a_c$ of the linear theory
 calculation and the phase-field simulation. Here the horizontal length
 of the simulation $L_x$ is taken to be equal to the wavelength of the
 perturbation $2\pi/k$ for every case. The number of grid points of the
 simulation is $N_x \times N_z = 128 \times 128$}.
 \label{t:ac}
\end{table}

Figure \ref{f:floquet} shows temporal variations of the interface
location obtained by the Floquet analysis results (\ref{floquet})
and the phase-field simulation results with critical $a_c$ values. 
We here follow the representation used in \cite{pjt}.
The initial discrepancies between the linear theory 
and the simulation remain large until one vibrating period $\Tv$.
However, later than that, the phase-field solutions agree
satisfactorily with the Floquet analysis result as shown
in figure \ref{f:floquet}.
In contrast, the discrepancies between the simulation 
by \cite{pjt} and the linear theory vanish within 
a quarter of the period of the vibration. 

Finally, we comment on the values of the mobility $\gamma$ and the
thickness of the interface $\epsilon$, which are adjustable numerical 
parameters of the phase-field simulation.
The mobility  $\gamma$ is determined as follows. 
The physical time scales of the various forces in (\ref{ns}) can be
dimensionally estimated for a given length scale $l_0$ as 
$t_g = (l_0 / g)^{1/2},\, 
t_\eta = l_0^2 \rho / \eta$ and 
$t_\sigma = (l_0^3 \rho / \sigma)^{1/2}$, which
are times scales of the gravity, viscosity and surface tension,
respectively.
The time scale associated with the mobility is in a similar way 
given as $t_\gamma = \gamma \rho$ by (\ref{ch}).
The  mobility value is determined by requiring
$t_\gamma < \min\{t_g, t_\eta, t_\sigma\}$ for a small length
scale, say $l_0 = \epsilon$. This means that the interface's 
relaxation to equilibrium occurs faster than other dynamics.
Here we have
$t_g = 4.29\times10^{-4}, 
t_\eta = 3.47\times10^{-5}$
and 
$t_\sigma = 3.34\times10^{-5}$s.
To have the condition $t_\gamma < \min\{t_g, t_\eta, t_\sigma\}$, 
we set $\gamma = 6.33 \times 10^{-8}$ kg$^{-1}$m$^3$s, which
yields $t_\gamma = 2.63\times 10^{-5}$s. 
That is, $t_\gamma / \min\{t_g, t_\eta, t_\sigma\} = 0.79$ here.
If this ratio is larger than $1.0$, our simulation does not agree
with the linear theory.
Here we take the time step to be $\Delta t = 2.50\times10^{-6}$s, which is 
about an order of magnitude smaller than $t_\gamma$.
The thickness $\epsilon$ is determined as follows.
We first fix the perturbation amplitude to $b = 3 \Delta z$ and then
do simulations with $\epsilon / \Delta z = 0.5, 1.0, 2.0, ..., 5.0$.
Good agreement with the $a_c$ of linear theory is obtained for $\epsilon /
\Delta z \le 3.0$. This leads us to choose $\epsilon=\Delta z$.

\section{Simulation beyond the linear regime} 
Having validated the phase-field approach in the linear regime,
we now move to nonlinear regimes.
Two cases are considered.
The first one concerns plume formation, which may eventually lead to
droplet ejection at long time. However, we do not follow the motion 
until the ejection or pinchoff occurs.
The second case is the period tripling state of
the two-dimensional Faraday waves.

\subsection{Plume formation}
\label{ss:pf}
Plume formation of the Faraday interfaces in two dimensions
was studied numerically by \cite{Poz00} using the vortex-sheet method, 
with which we compare the phase-field simulation.
The presence of a plume implies that the interface profile 
becomes a multivalued function of the horizontal coordinate.
Such situations can be numerically handled by the phase-field method 
without \textit{ad hoc} adjustments. 

Here the comparison is qualitative. 
There are three reasons for this: 
the difference in Atwood number $A = {(\rb - \rt)}/{(\rb + \rt)}$,
the difference in the boundary conditions,
and with or without viscosity.
Regarding the Atwood number difference, 
the phase-field method with the discretization described
in section \ref{ss:nm} is not able to handle high Atwood
number $A$. Our simulation works for at most 
$A \simeq 0.40$ in practice (this issue is discussed in
the last \S \ref{s:discon}).
In contrast, the vortex-sheet calculation by \cite{Poz00} can 
cope with up to unit Atwood number. 
Concerning the difference in the boundary conditions,
there is no rigid boundary in the vertical direction
in their vortex-sheet calculation. However, in our case fluids are contained
between two rigid walls.
The third reason is that the vortex-sheet calculation is inviscid, while the phase-field method
is viscous.

\begin{figure} 
 \centerline{\includegraphics[height=2in,width=3in]{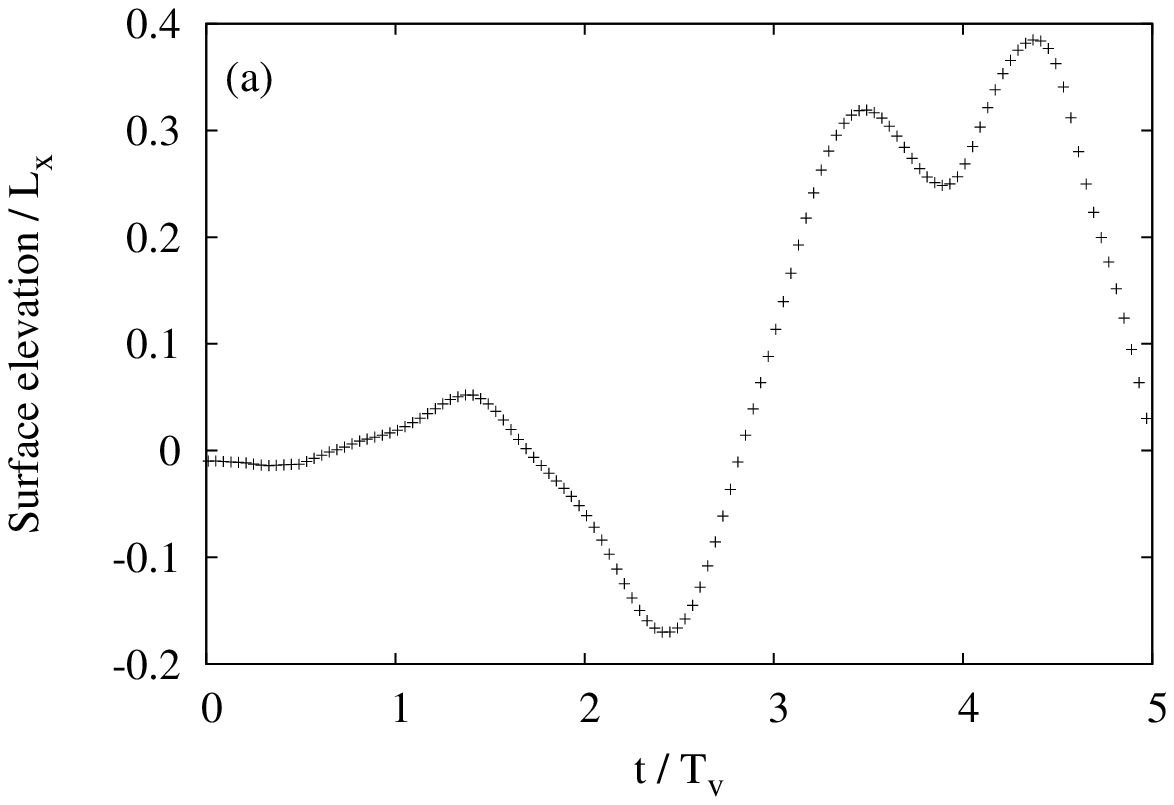}} 
 \centerline{\includegraphics[scale=0.8]{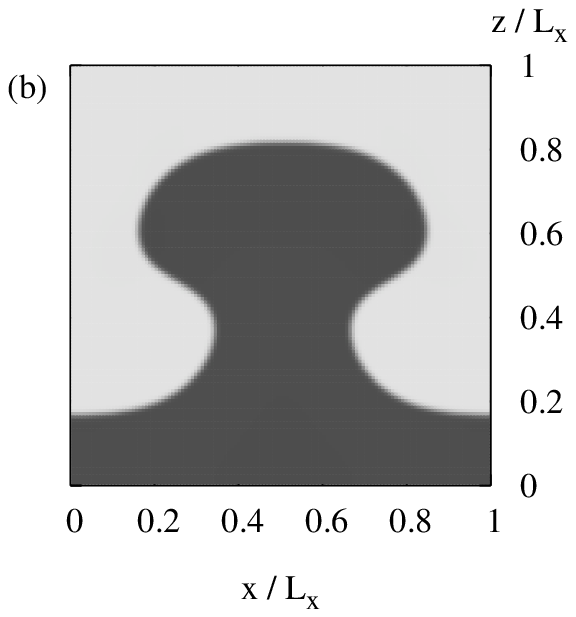}
             \includegraphics[scale=0.8]{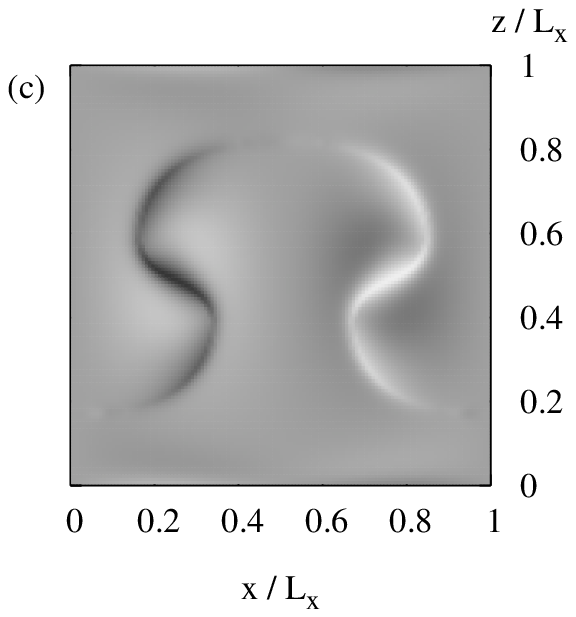}}
 \centerline{\includegraphics[scale=0.8]{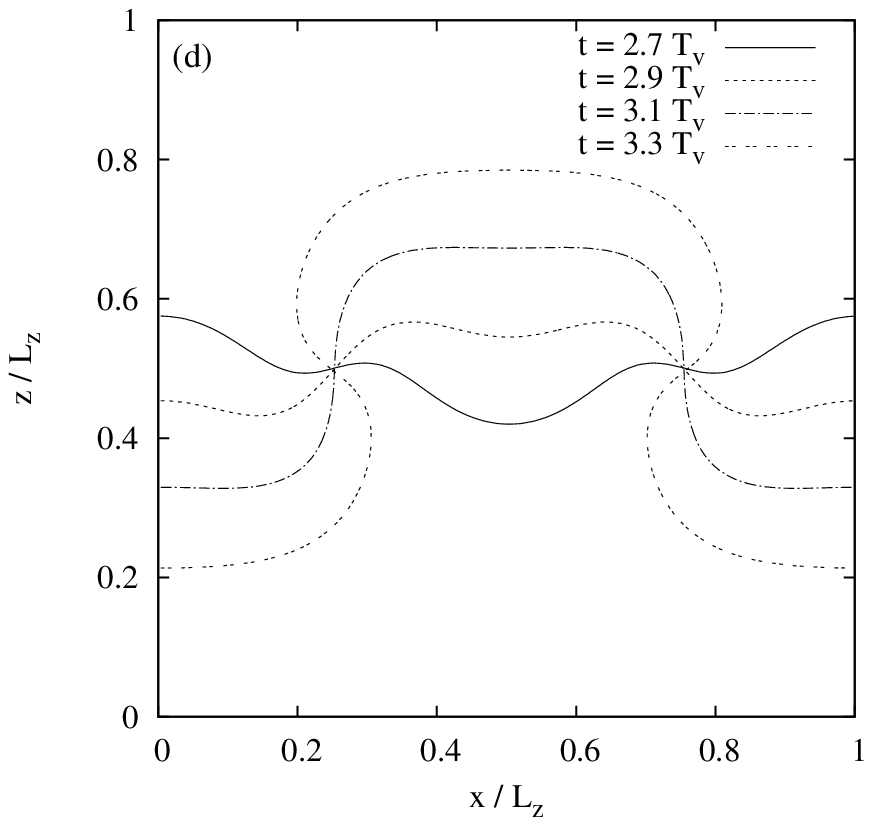}
             \includegraphics[scale=0.8]{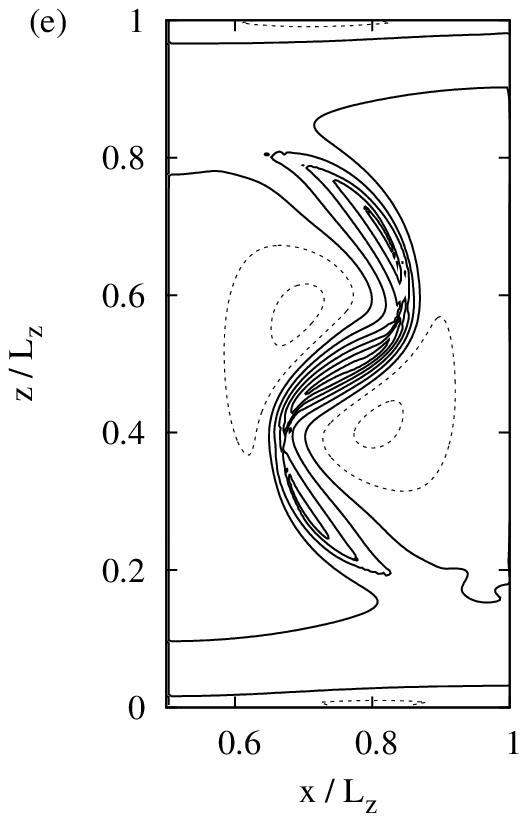}}
 \caption{Faraday interface in a nonlinear regime, showing 
an overturning behaviour in the phase-field simulation.
 (a) Time variation of the interface position at $x = L_x / 2$. The
 interface position is defined from the phase field $\phi$
 as described in section \ref{s:linear}.  
 (b) Grey-scale coded phase field $\phi$ at time $t = 3.5 \Tv$.
(c) Grey-scale coded vorticity field at the same instant.
(d) Interface profiles at four instants.
(e) Contours of the vorticity at $t = 3.5 \Tv$. Contour values are
from -20.0 to 20.0 by 4.0. The solid and dashed lines correspond to
non-negative and negative vorticity values.
} 
 \label{f:ot}
\end{figure}

In \cite{Poz00}, the overturning interface corresponding to 
formation of a plume at $A = 0.65$ and $1.0$ is numerically 
studied.
We observe a similar overturning behaviour in the phase-field simulation even at as low as $A = 0.10$, which we show 
in figure \ref{f:ot}.
The parameters of the phase simulation are: 
$L_x = 1.0 {\rm m}, L_z = 1.0 {\rm m}, a = 9.0 \times 10^{-2} {\rm m
s^{-2}}, \omega = 2.6728 \times 10^{-1}{\rm s^{-1}}, \rb = 1.0 \times 10^{3} {\rm kg m^{-3}}, \rt
= 8.1818 \times 10^{2}{\rm kg m^{-3}}, \etat = 0.2 {\rm Pa s}
, \etab = 0.2 {\rm Pa s},
\sigma = 7.2 \times 10^{-2}{\rm N m^{-1}}, g = 0.0, $
$\Lambda = 3\sigma / (2\sqrt{2} \epsilon), \gamma = 5.61 \times
10^{-4}{\rm m^{3} kg^{-1} s}$ and $ \epsilon = \Delta z$. 
The force $\bm{G}$ in (\ref{ns}) is here $\bm{G} = a\cos\omega t$,
oscillating around zero.
The initial interface is given by (\ref{e:tanh}) with $b = 0.01 {\rm m}$. 
The time step and the number of grid points are $\Delta t = 1.18
\times10^{-3}$
and $N_x \times N_z = 256 \times 256$. 
Here we take the same parameters $L_x, \omega, \rb, b, \sigma$ and 
the same vibration force $\bm{G}$ setting (oscillating around the
origin) as \cite{Poz00}. 
The differences between our simulations and theirs are the Atwood number,
the vibration amplitude value (ours are nine times larger),
the presence of the top and bottom rigid walls 
and the presence of viscosity.

The time variation of the interface location at a fixed horizontal 
point is shown in figure \ref{f:ot}(a).
The interface location first exhibits a rise and then a fall
 (here at about $1.5\Tv$ and $2.5 \Tv$, respectively, where $\Tv$ is the
 vibration period);
then it grows steeply (at around $t = 3.0 \Tv$) 
and reaches a high plateau (between $t = 3.4 \Tv$ and $t = 4.4 \Tv$)
with a small oscillation.
This variation is  similar to 
that with $A = 0.65$ obtained with the vortex-sheet method 
(figure 11(a) of \cite{Poz00}).
In our simulation, the interface location later than $t = 5\Tv$ (continuation of
figure\ref{f:ot}(a) but not shown) goes downwards and shows again a
plateau with a small oscillation. Then it goes upwards and finally comes back
to zero around $t=8\Tv$. This later variation cannot be compared with
the vortex-sheet method result of \cite{Poz00} because their calculation
is stopped before $t=5\Tv$.

The phase field $\phi$ at $t = 3.5 \Tv$ is shown
in figure \ref{f:ot}(b), which clearly shows that the interface turns
over and becomes
a multivalued function of the horizontal coordinate $x$. 
Qualitatively similar interface structures, called plumes,
are shown in figure 11(b) of \cite{Poz00}. 
To check the boundary effect, we also do the phase-field simulation with aspect ratio 2 ($L_x =
2L_z$). 
The plume shape and its time evolution do not change qualitatively,
indicating that the effect of the vertical boundary is small on the
plume formation.
This agreement suggests that the phase-field simulation
is valid, at least qualitatively, in the overturning regime 
although it cannot handle cases with high Atwood number.

In figure \ref{f:ot}(c) and (e), we plot the vorticity field 
$\Omega = \partial_z u - \partial_x w$ at the plume state.
Indeed much of the vorticity concentrates inside the interface 
region. The vorticity along the interface in the right half domain
has three peaks as seen in the contour plot of figure \ref{f:ot}(e).
In contrast, in the simulation of \cite{Poz00}, the strength of the vortex
sheet has at most two peaks in the developed plume state. 
In the phase-field simulation we observe
that the vorticity outside of the interface reaches about 
40\% of the maximum vorticity in the interface region.
In addition, thin vortex layers are attached near the top and bottom walls.
In these layers, the vorticity is about 10\% of the maximum in the 
interface region.

We now comment on the choice of the resolution $N_x \times N_z = 256
\times 256$ used here. 
We observe that the squared modulus of the Fourier coefficients
$|\hat{\phi}(k, z, t)|^2$ decreases exponentially for large $k$ such as
$|\hat{\phi}(k, z, t)|^2 \propto \exp[-\delta(z, t) k]$. We measure this
factor $\delta(z, t)$ and require $\max_{z, t}[\delta (z, t)] \ge 2
\Delta x = 2 L_x / N_x$. We take the smallest $N_x$ in powers of 
2 satisfying this relation. We then set $\Delta z$ to about the same
value of $\Delta x$.

\subsection{Period tripling}
\label{s:pt}
The period tripling state of Faraday waves,
in which the period of the standing wave becomes 
three times the usual subharmonic period (which is in turn twice the
vibration period),
is found in the quasi-two-dimensional experiment
by \cite{jian98} (see also \cite{ps}) and 
in the three-dimensional axisymmetric experiment
by \cite{jian98} and also by \cite{dh08}. 
In the two-dimensional vortex-sheet simulation of \cite{Poz00},
period tripling is also observed. 
From their numerical result, \cite{Poz00} conclude
that the period tripling is caused by
the nonlinearity of the irrotational flow. 
We here aim at reproducing the period tripling
state with the phase-field modelling within our feasible
Atwood number range $A < 0.40$, although the above experiments
and the vortex-sheet simulation are conducted in the case 
$A \simeq 1.0$.

\begin{figure} 
 \centerline{\includegraphics[height=2in,width=3in]{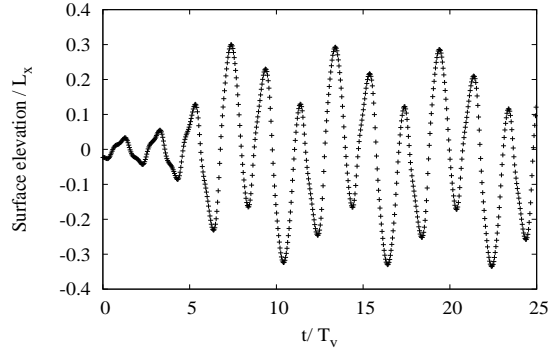}}
 \caption{Temporal variation of the interface position at $x = L_x / 2$,
 showing the period tripling. Here $\Tv$ is the period of the vibration forcing.
 }
 \label{f:pt_fixed_x}
\end{figure}

\begin{figure}  
 \centerline{\includegraphics[scale=0.7]{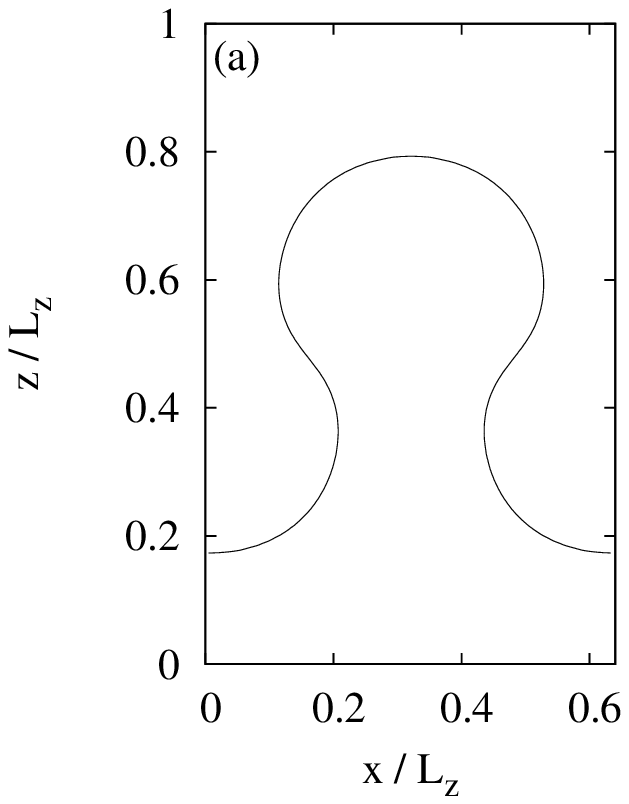}  
             \includegraphics[scale=0.7]{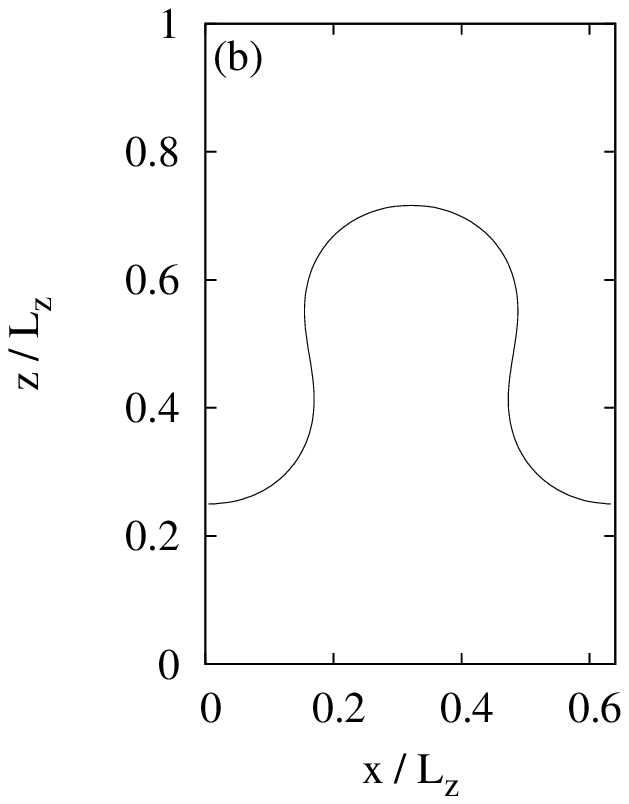}
             \includegraphics[scale=0.7]{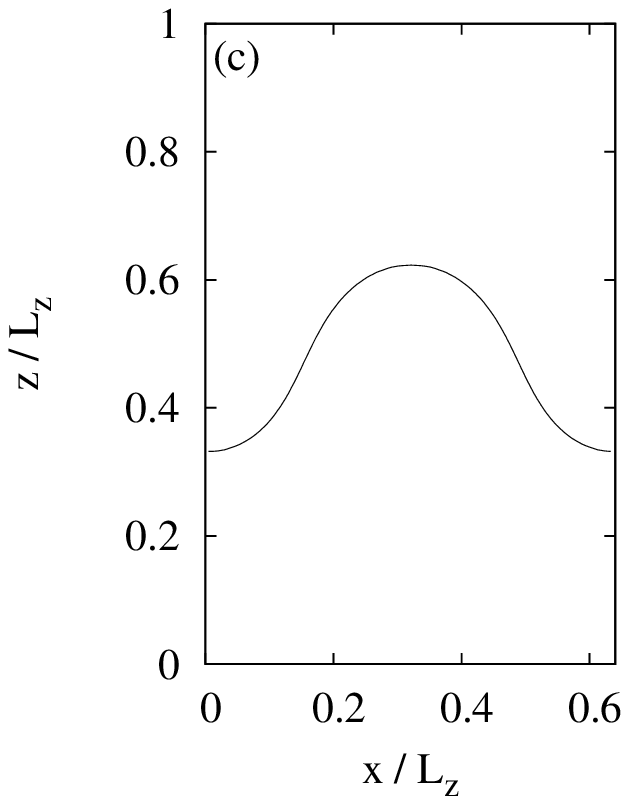}}
 \centerline{\includegraphics[scale=0.7]{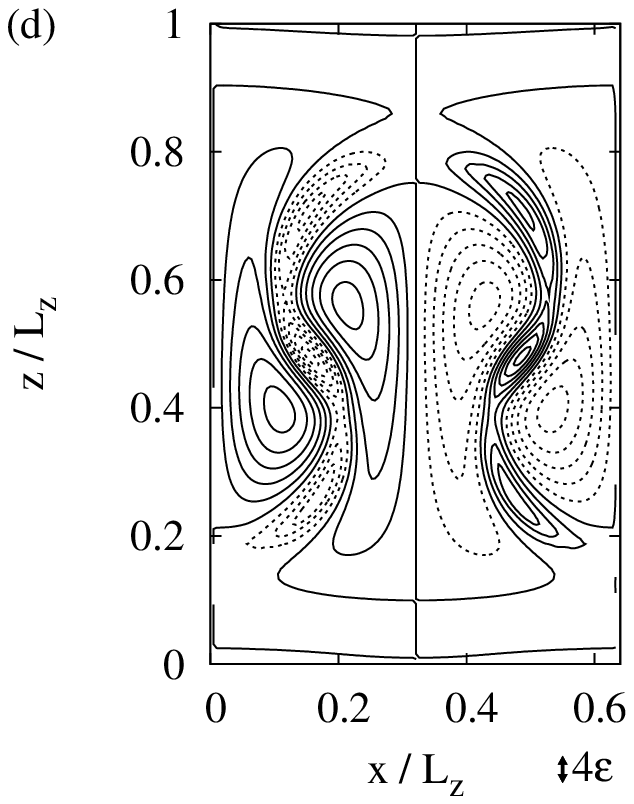}  
             \includegraphics[scale=0.7]{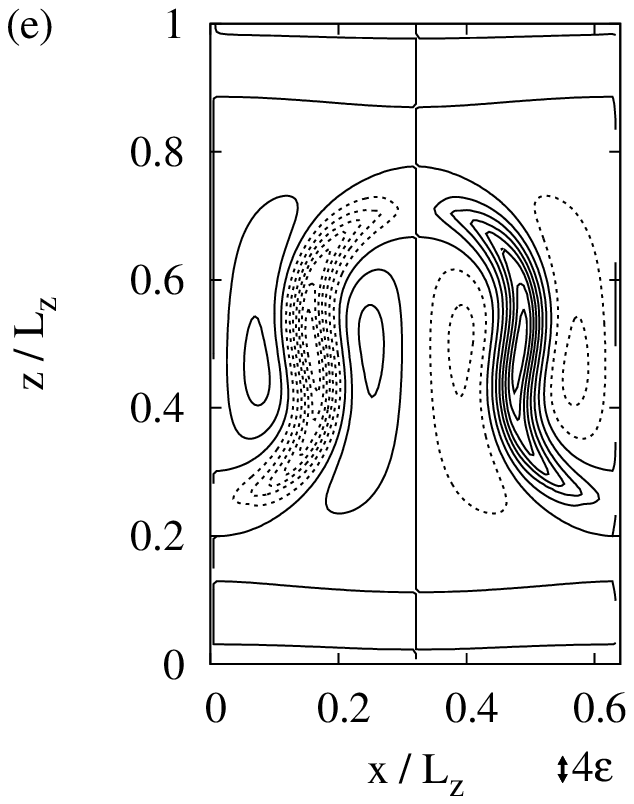}
             \includegraphics[scale=0.7]{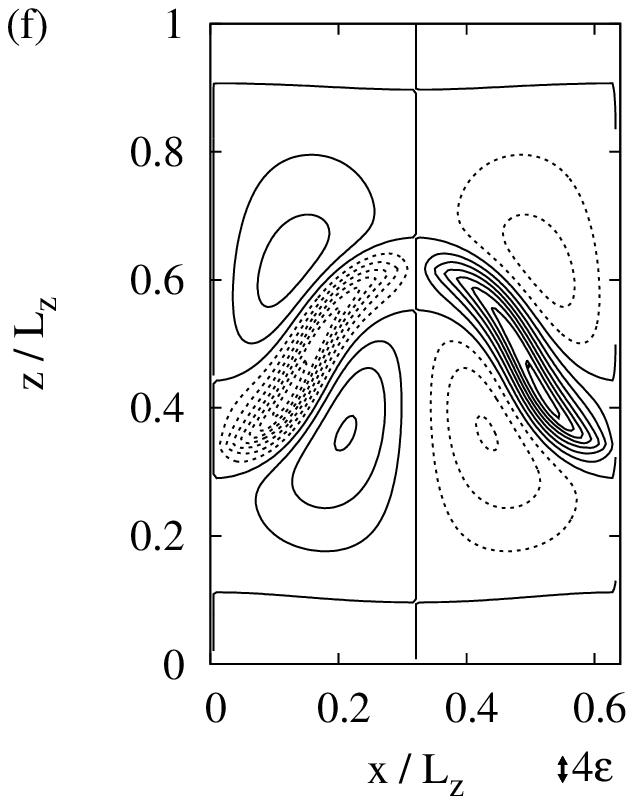}} 
 \caption{Interface profiles corresponding to the three peaks of the period tripling
 state (a) $t = 19.4 \Tv$; (b) $t = 21.4 \Tv$; (c) $t = 23.4 \Tv$. They
 correspond to the last three peaks shown in figure \ref{f:pt_fixed_x}. 
 Vorticity contours (solid and dotted lines correspond to non-negative and
 negative values, respectively)
at the same instances:
 (d) $t = 19.4 \Tv$, the contour values are from $-5.0$ to $5.0$ by $1.0$;
 (e) $t = 21.4 \Tv$, the values are from $-7.0$ to $7.0$ by $1.0$;
 (f) $t = 23.4 \Tv$, the values are the same as (e).
  }
 \label{f:3modes}
\end{figure}

By searching the parameter space, which will be discussed later, 
indeed we observe the period tripling state at low Atwood number $A=0.11$, 
as depicted in figure \ref{f:pt_fixed_x}. The interface shapes at the 
maxima are shown in figure \ref{f:3modes}, some of which actually
differ from the previous observations.
In the experiment by \cite{jian98}, the interface shapes at the three peaks 
in the tripling state are classified 
as sharp crest (called mode A by them), flat or dimpled crest (mode B) and round crest
(mode C), respectively.
The corresponding shapes of the phase-field simulation 
shown in figure \ref{f:3modes}(a--c)
are qualitatively different except for the round crest
(figure \ref{f:3modes}(c)).
In particular, the hourglass plume shown in figure \ref{f:3modes}(a) and
waistless plume in figure \ref{f:3modes}(b) are different
from the sharp crest (mode A) and the flat or dimpled crest (mode B). 
These sharp and flat crests are associated with wave breaking in \cite{jian98} 
but not necessarily so in \cite{dh08}. In the phase-field simulation
with this parameter set, wave breaking does not occur.
In the period tripling state observed numerically by \cite{Poz00},
plume-type interfaces are not observed.

The vorticity contours at the three peaks are shown in figure
\ref{f:3modes}(d--f). The number of vortex peaks inside the 
interface region is three for the hourglass plume and one for the
waistless plume and the round crest. 
Outside of the interface region the vorticity is weaker but not zero.

\begin{figure} 
 \centerline{\includegraphics[height=2in,width=3in]{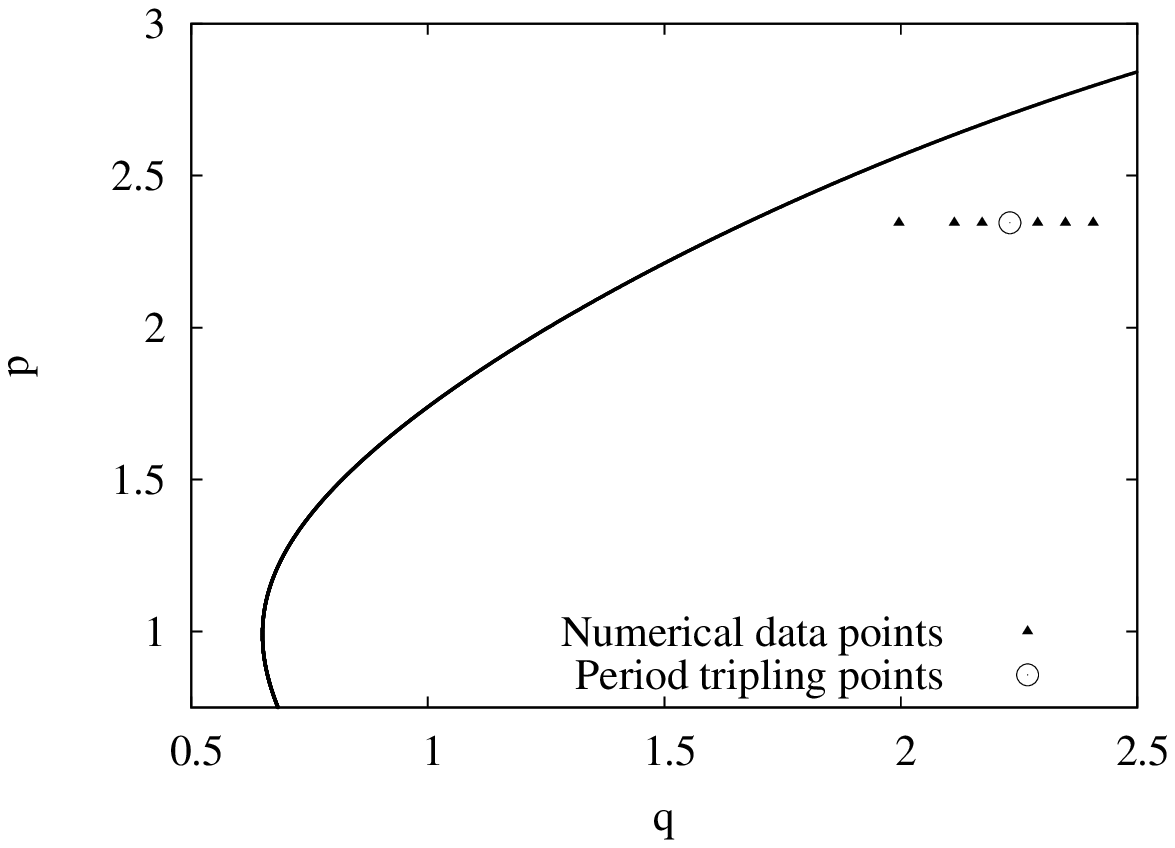}}
\caption{The $p, q$ parameter space for seeking the period tripling
 regime. Here $p = \{2\Omega(k)/\omega\}^2$ is the detuning
 parameter and $q = 2Ak a/\omega^2$ is the forcing parameter,
where $\Omega^2(k) = k g [ A + \sigma k^2 /\{g(\rb + \rt)\}] $.
The points are $q = 2.00, \,2.11, \,2.17,\, 2.23,\, 2.29,\, 2.35,\, 2.41$ and constant $p = 2.35$.  
The curve corresponds to the neutral stability line obtained by 
the linear analysis of \cite{kt94} (the method described in \S
 \ref{s:linear} is used).
}
 \label{f:pqdiagram}
\end{figure}

To find the parameters of the period tripling state in practice,
we use as a guide the phase diagram made by \cite{jian98} in 
the space of the detuning parameter $p$ and forcing parameter $q$.
We search this $p,q$-space horizontally (constant $p$) by going away 
from the neutral curve as shown in figure \ref{f:pqdiagram}. 
We take seven parameter sets along this line. 
The period tripling state is found at the middle point.
In the phase diagram in \cite{jian98}, the period tripling state 
extends for the large $q$ region. But our case is very different.
The parameters of the tripling state are: 
$L_x          = 1.46 \times 10^{-4}$ m,
$L_z          =  2.31 \times 10^{-4}$ m,
$a/g          = 9.5$,
$\omega       = 2\pi \times 10^2$ s$^{-1}$, 
$\rho_{\rm b} = 5.19933 \times 10^{2}$ kg m$^{-3}$,
$\rho_{\rm t} = 4.15667 \times 10^{2}$ kg m$^{-3}$,
$\eta_{\rm b} = 3.908   \times 10^{-5}$ Pa s,
$\eta_{\rm t} = 3.124   \times 10^{-5}$ Pa s,
$\sigma       = 2.181   \times 10^{-6}$ N m$^{-1}$,
$g            = 9.8066                $ m s$^{-2}$,
$\epsilon = \Delta z = 1.80 \times 10^{-5}$ m, 
$\gamma = 6.33 \times 10^{-8} \,{\rm m^3 kg^{-1}s}$,
$\Lambda = 3\sigma / (2\sqrt{2} \epsilon)$
and
$N_x \times N_z = 128 \times 128$. 
The spatial resolution is
determined in the same way as described in subsection \ref{ss:pf}.

Now we discuss the behaviours at other parameter points shown in figure 
\ref{f:pqdiagram}.
For three $q$ values smaller than the period tripling value, 
we observe that the period of the standing wave remains twice the 
vibration period (the subharmonic period).
The interface shape at the maximum and minimum surface elevations 
are hourglass-plume type.
In other words, the temporal symmetry is kept as we approach
the period tripling state
from the left while in the experiment and numerical simulation by 
\cite{jian96} the symmetry is broken before reaching the tripling 
state.
This discrepancy may be due to the difference in the Atwood number
or the difference in the dissipation caused by the sidewalls.
For the next larger $q$ parameter than that of the tripling state, 
we observe that the period of the standing wave returns to 
the subharmonic period ($2\Tv$).
We do not have an explanation for this.
The interface shape at the maximum and minimum elevation is also
of the hourglass-plume type. 
For the two largest $q$ parameters that we calculate, 
the oscillation becomes close to quasi-periodic. 
The interface shapes at relative maximum
elevations in this parameter range take the three forms shown in figure
\ref{f:3modes}(a-c) almost randomly.

\section{Discussion and concluding remarks}
\label{s:discon}

We have applied the phase-field modelling of the binary fluids
with the Cahn-Hilliard equation due to \cite{Jacqmin} to 
numerical simulations of the Faraday wave problem in two spatial 
dimensions. 
Here we have solved the Navier-Stokes equations for both the 
top and bottom fluids with rigid boundary conditions
on the top and bottom walls and with periodic boundary condition
for the side walls.
Validation of this phase-field simulation is checked quantitatively
in the linear regime and qualitatively in the nonlinear regime.

In the linear regime, our simulation agrees quantitatively 
well with the Floquet analysis by \cite{kt94} in three different 
branches, which is the benchmark test proposed in \cite{pjt} for 
the Faraday waves.

In the nonlinear regime, we are not able to validate the present 
simulation against experimental results unlike \cite{pjt}.
Nevertheless we have considered two cases. 
Both cases involve plume formation where the two-fluid interface becomes
a multivalued function of a horizontal coordinate.
Such situations can be a good testing ground for the phase-field
method since the method does not break down when the interface turns
over. 

The nonlinear case we considered first is the bursting plume
formation which was studied by \cite{Poz00} with the vortex-sheet
method.
Their system setting (the boundary conditions, the dissipation and
close-to-unity Atwood number, especially) is very different from the present
binary fluid setting. Consequently the comparison is necessarily
qualitative. The phase-field method with the numerical scheme
described in section 3 works only in the Atwood number range $A < 0.40$.
Nevertheless we found for very low Atwood number $A = 0.10$ that a similar 
plume is formed and the temporal variation of the interface agrees
qualitatively with that in \cite{Poz00}.

The second nonlinear case is the period tripling of the two-dimensional
Faraday waves first found experimentally by \cite{jian98} and numerically
by \cite{Poz00}.
Again, in spite of the rather big system differences with these studies, 
we have observed the period tripling state in 
the phase-field simulation with low Atwood number $A = 0.11$. 
This result adds further evidence for the robustness of the period
tripling as discussed in \cite{jian98}.
More detailed numerical study of the tripling state is currently 
under way and will be reported elsewhere.
It would be interesting to perform a laboratory experiment with the same
physical parameters as in \S \ref{s:pt} to check whether or not the period
tripling is observed. 

%%%%%%%%%%%%%%%%%%%%%%%%%%%%%%%

Concerning the limitation on the Atwood number $A < 0.40$ in the present
phase-field simulation, we do not have an explanation for why it is so.
For $A \ge 0.40$, we observe that, if we put a larger number of grid 
points inside the interface region, we can postpone the breakdown of the
simulation until somewhat later (we can continue the simulation for a larger number of time steps). 
However, increasing the number of grid points there
may cause a new problem. Although \cite{Jacqmincontact}
successfully reaches Atwood number $0.98$, he reports that the interface 
becomes much wider than in reality. It is not clear to us for the present 
whether or not we can overcome these problems  by improving the temporal 
and spatial discretizations of the Cahn-Hilliard and Navier-Stokes equations.

Finally, we comment on extension of the phase-field method to three-dimensional Faraday waves. It is straightforward. 
We believe that the three-dimensional phase-field method, once extended, 
can be complementary to the simulation method by \cite{pjt} 
since the two methods are different.
There is an interesting instance of three-dimensional Faraday waves, where
the interface becomes multivalued. That is the propagating
solitary state observed by \cite{LAF}.
Our simulation code can easily study such a state.

We acknowledge the support by the Grant-in-Aid for the Global COE Program
"The Next Generation of Physics, Spun from Universality and Emergence"
from the MEXT of Japan and by the Grant-in-Aid for Young Scientists (B)
No. 2174290 from the JSPS. We thank the anonymous referees for
constructive criticisms.

%%%%%%%%%%%%%%%%%%%%%%%%%%%%%%%%%%%%%%%%%%%%%

\end{document}